\documentclass[12pt]{article}
\usepackage{graphics}
\usepackage{graphicx, subfigure}
\usepackage{amssymb}
\usepackage{amsmath}

\newcommand{\rf}[1]{(\ref{#1})}
\newcommand{\beq}{\begin{equation}}
\newcommand{\eeq}{\end{equation}}
\newcommand{\bea}{\begin{eqnarray}}
\newcommand{\eea}{\end{eqnarray}}

\newcommand{\e}{\mbox{e}}

\newcommand{\g}{\gamma}

\renewcommand{\b}{\beta}

\newcommand{\m}{\mu}

\newcommand{\Del}{\Delta}

\newcommand{\ra}{\rangle}
\newcommand{\la}{\langle}
\newcommand{\prt}{\partial}

\newcommand{\cD}{{\cal D}}

\begin{document}

\begin{center}
\vspace{24pt}
{ \large \bf The toroidal Hausdorff dimension of 2d  Euclidean 
quantum gravity}

\vspace{30pt}

{\sl J. Ambj\o rn}$\,^{a,b}$
and 
{\sl T. Budd}$\,^{a}$

\vspace{48pt}
{\footnotesize

$^a$~The Niels Bohr Institute, Copenhagen University\\
Blegdamsvej 17, DK-2100 Copenhagen \O , Denmark.\\
email: ambjorn@nbi.dk, budd@nbi.dk\\

\vspace{10pt}

$^b$~Institute for Mathematics, Astrophysics and Particle Physics (IMAPP)\\ 
Radbaud University Nijmegen, Heyendaalseweg 135,
6525 AJ, Nijmegen, The Netherlands

}
\vspace{96pt}
\end{center}


\begin{center}
{\bf Abstract}
\end{center}

\noindent 
The lengths of shortest non-contractible loops are studied numerically in 2d Euclidean quantum gravity on a torus coupled to conformal field theories with central charge less than one.
We find that the distribution of these geodesic lengths displays a scaling in agreement with a Hausdorff dimension given by the formula of Y. Watabiki.

\vspace{12pt}
\noindent

\vspace{24pt}

\newpage

\section{Introduction}\label{intro}

The path integral plays an important role in quantum
mechanics and quantum field theory. A feature
of the path integral is that a ``generic'' path $X_i(t)$, 
$i=1,\ldots,d$, is fractal
with Hausdorff dimension $D_H=2$. This is true irrespective of
the dimension $d$ of space. The word ``generic'' refers to 
the path integral when rotated to imaginary time. When this
is done one has a measure on the set of parametrized 
continuous paths, the Wiener measure, and with respect to
this measure  a randomly chosen continuous path 
has Hausdorff dimension 2.  While the free non-relativistic 
particle is described by the path integral over parametrized 
paths in $d$ dimensions,  
the path integral for the free {\it relativistic} particle can  
be defined by considering  paths in  $D=d+1$ dimensional spacetime 
and use the Brink--Di Vechhia--Howe  action for the particle,
where one includes an intrinsic metric on the world line
as a dynamical variable. The path integral
can thus be viewed as the path integral of 
one-dimensional quantum gravity coupled 
to $D$ Gaussian fields $X_\m$, the coordinates
of space-time (see \cite{book} for a review). Still, in this approach the 
Hausdorff  dimension of the ensemble of paths in $D$ dimensions
is two, while the dimension of the ``intrinsic'' parameter space 
trivially is one.

The generalization of the Brink--Di Vecchia--Howe action to 
strings leads to the Polyakov path integral.
In this theory we have an intrinsic two-dimensional metric 
on the world sheet and coupled to this $D$ Gaussian fields $X_\m$.
In this case we formally define the path integral as follows
\beq\label{1.1}
Z_V = \int \cD [g]_V \int \cD [X_m]_{{\rm cm}} \;
\e^{-\int d^2\xi \sqrt{g} \, g^{ab}\prt_a X_\m \prt_b X_\m},
\eeq
where the integration is over two-dimensional geometries 
with fixed volume  $V$ and where the subscript ``cm'' signifies that
the center of mass of the string 
is fixed, i.e.
\beq\label{1.2}
\int d^2\xi \sqrt{g(\xi)}=V,~~~~\int d^2\xi X_\m(\xi) =0.
\eeq
We note that one can in principle perform the Gaussian 
integration over the fields $X_\m$ and one finds
\beq\label{1.1a}
Z_V \propto  \int \cD [g]_V \left( \det (-\Del_g) \right)^{-D/2},
\eeq
where $\Del_g$ is the Laplace operator in the geometry 
defined by the metric $g_{ab}$. Note that this expression allows 
us an to treat also non-integer values for $D$. 
The so-called {\it extrinsic} Hausdorff dimension $D_H$ is 
then defined by 
\beq\label{1.3}
\la X^2\ra_V \sim V^{2/D_H}
\eeq
where 
\beq\label{1.4}
\la X^2\ra_V = \frac{1}{ V Z_V} 
\int\! \cD [g]_V\!\! \int \!\cD [X_m]_{{\rm cm}} \;
\e^{-\int d^2\xi \sqrt{g} \, g^{ab}\prt_a X_\m \prt_b X_\m}\; 
\int d^2\xi \sqrt{g}\; X_\m^2(\xi)
.
\eeq
This definition, when applied to the particle where the integration
is over $d\xi$ rather than $d^2\xi$, leads to $D_H=2$ as mentioned 
above. In the case of string theory one encounters tachyons 
when $D > 1$, and the corresponding Liouville theory is ill 
defined. For $D \leq 1$ one obtains \cite{spectrald}
\beq\label{1.5}
\la X^2 \ra_V \sim \log^2 V ~~(D=1),~~~~~\la X^2 \ra_V \sim \log V~~(D<1),
\eeq
where there result  for $D <1$ is obtained by analytic continuation
as mentioned above,
by formally considering the Gaussian theory as a conformal field theory 
with central charge $c=D \leq 1$.
Thus $D_H = \infty$ for non-critical string theory.

However, contrary to the particle case, we can define a 
non-trivial {\it intrinsic} Hausdorff dimension. This 
dimension is natural and of interest when we view non-critical 
string theory as two-dimensional (Euclidean) quantum gravity 
coupled to some conformal theory with central charge $c<1$.
For such a conformal field theory one does not in general have a 
natural definition of $D_H$ which refers explicitly to the Gaussian 
fields $X_\m$, but we can, by analogy with \rf{1.3}, 
define the {\it intrinsic} Hausdorff dimension as 
\beq\label{1.6}
\la r^2\ra_V \sim V^{2/d_h}.   
\eeq
The average is defined with respect to the  partition function $Z_V$: 
\beq\label{1.7}
Z_V= \int \cD [g]_V Z_V (g,c),
\eeq
where $Z(g,c)$ denotes the partition function for the  
conformal field theory with central charge $c$ we consider, 
in the ``background geometry'' defined by the 2d metric $g_{ab}(\xi)$.
The functional integration in \rf{1.7} is over two-dimensional 
geometries with volume $V$, as defined in eq.\ \rf{1.2}, and the 
average in \rf{1.6} is now
\beq\label{1.8}
\la r^2 \ra_V = \frac{1}{ 2 V^2 Z_V} \int \cD [g] Z_V(g,c) 
\int \!d^2\xi \!\int \!d^2\xi' 
\sqrt{g(\xi)} \sqrt{g(\xi')} \; r^2_g(\xi,\xi')  
\eeq
where $r_g(\xi,\xi')$ denotes the geodesic distance between $\xi$ and $\xi'$
in the geometry defined by $g_{ab}(\xi)$. 

A remarkable formula for $d_h$ was derived by Y. Watabiki \cite{watabiki},
using Liouville theory and the heat kernel expansion:
\beq\label{1.9}
d_h^W(c) = 2 \,\frac{\sqrt{49-c}+\sqrt{25-c}}{\sqrt{25-c}+\sqrt{1-c}}.
\eeq
This formula is not the only one proposed. Already in the original 
articles where quantum Liouville theory was defined an alternative 
formula was suggested \cite{dk}
\beq\label{1.10}
d_h'= -\frac{2}{\g}, ~~~\g= \frac{c-1-\sqrt{(1-c)(25-c)}}{12}.
\eeq
The two formulas agree for $c=0$ where $d_h^W=d_h'=4$.
However, they have a quite different behavior for $c\to -\infty$
where $d_h'\to 0$ while $d_h^W \to 2$. Formally one expects $d_h \to 2$
since this is the limit where the quantum Liouville theory should 
behave semiclassically and matter and geometry should 
be only weakly coupled, and this is indeed the behavior of $d_h^W$.
On the other hand it is quite difficult to provide any geometric 
interpretation of $d_h' \to 0$. The difference 
between the two predictions becomes even more pronounced when 
we consider the region $0<c<1$. For $c \to 1$ $d_h' \to \infty$
while $d_h^W \to 2+2\sqrt{2} =4.83$.

That the value $d_h=4$ is correct for $c=0$ has been proven
in \cite{transferm,aw}. 
The present understanding is that $d_h'$ does not reflect the 
behavior \rf{1.6} with $r$ being the geodesic distance when $c \neq 0$.
Rather it reflects the behavior of some structures related to matter.
This is nicely exemplified for $c=1/2$. The continuum  $c=1/2$
theory in flat spacetime can be obtained as the scaling limit 
of the Ising model defined on a regular 2d lattice. The Ising model
has a critical temperature (or coupling constant)  
and a second order phase transition where 
one can define the $c=1/2$ conformal field theory. Similarly 
the continuum 2d quantum gravity theory coupled to a $c=1/2$ conformal
field theory can be obtained as the scaling limit of an Ising model
coupled to so-called ``dynamical'' triangulations \cite{kazakov}, 
i.e.\ rather than
considering the Ising model on a  fixed regular lattice, one considers 
the Ising model on  lattices with fixed 2d topology, but otherwise
random connectivity, and  sums over all such random lattices.
If we compare with formula \rf{1.7} the number of vertices can 
(scaled suitably) be identified with $V$, the sum over geometries
can be identified with the sum over lattices with different
connectivities (and it can all be made quite precise using the 
formalism of dynamical triangulations (DT), see \cite{book} for a review).
The Ising model on this so-called annealed average of lattices still has a 
critical point and a corresponding phase transition where the 
continuum limit can be obtained. The critical exponents 
one finds in this limit are precised the KPZ exponents of a $c=1/2$
theory coupled to 2d quantum gravity. For $c=1/2$ one obtains 
$d_h'=6$. The seminal work of Kawai and Ishibashi \cite{ki} shows 
that one obtains $d_h=6$ in the DT Ising model, not by using the geodesic 
distance (for instant the shortest link distance between two vertices),
but by counting the number of spin boundaries separating the two vertices.
Of course it might be that for large volumes the average number of spin
boundaries separating two vertices is proportional to the average number of 
links separating the vertices. However, computer simulations do not 
support this, but suggest that the geometric Hausdorff dimension is 
closer to the value $d_h^W(c=1/2)= 4.21$ suggested by the Watabiki formula. 

Almost the same story can be repeated for $c=-2$ theories. In this 
case $d_h^W=3.562$ while $d'_h =2$. Analytically it is possible 
to obtain the $d'_h$ value \cite{aajk} from a specific 
discretized theory, the $O(n)$ model coupled to DT, where 
the value $n=-2$ corresponds to  $c=-2$, but again this value 
is obtained by using in \rf{1.6} an $r$ with no direct 
relation to the geometric distances on DT lattices but rather to 
spin boundaries in the $O(n)$ model. On the other hand it is formally 
possible to realize the $c=-2$ model on dynamical triangulations
if one uses the Polyakov string partition function \rf{1.1a} analytically 
continued to $D=-2$ fields. Remarkably, it is possible to 
find an algorithm which generates the triangulations recursively 
with the correct weight in this case \cite{kas}, allowing one 
to numerically determine the geometric $d_h$ for very large 
triangulations \cite{dminustwo}. One finds perfect agreement 
with the Watabiki formula. Recently, measurements of the 
geometric $d_h$ have been performed for large negative $c$, based 
on Monte Carlo simulations using 
the partition function \rf{1.1a}, again analytically continued 
to large negative $D$. One observes agreement with the Watabiki 
formula  \cite{ab3}. Also, one observes (at a qualitative,
visual level \cite{ab3}) that the typical geometries generated in the 
computer simulations become less fractal for large negative $D$ in
accordance with the expectation that smooth geometries should 
dominate in the $c=-\infty$ limit in the Liouville theory. This 
is probably the reason it is possible at all to determine 
$d_h$ numerically for large negative $D$ using \rf{1.1a}, since
the presence of the determinant for numerical reasons constraints
the size of the triangulations (except for $D=-2$).

It is thus fair to say there is good ``empirical'' evidence 
that \rf{1.9} is the geometric Hausdorff dimension for 
$c\leq 0$. The purpose of this article is to test numerically whether \rf{1.9} is valid also in the range $0 <c <1$. Why is this needed? One should keep 
in mind that the analytic arguments which led to \rf{1.9}
are not rigorous arguments, as emphasized by Watabiki himself
\cite{watabiki}. They rely on the interchangeability of certain 
limits when  averaging over geometries and averaging over diffusion-paths,
which is not necessarily true. It might be true for $c \leq 0$
but not for $c>0$ where the average geometry could  become 
more fractal. In fact there {\it is} distinction between the 
regions $c <0$ and $c>0$. For minimal conformal theories
with $c>0$ coupled to 2d gravity the dominant infrared 
coupling constant is the cosmological coupling constant,
while for $c<0$ one has primary fields of negative dimensions
and the coupling constant related to the primary 
operator of the most negative dimension is expected 
to dominate the infrared in an effective field theory.
We do not presently understand how such a dominance is 
transferred to a change of $d_h$, but if that is the case
one could have a scenario where $d_h$ varies for $c<0$ (and 
agrees with \rf{1.9}), while it stays 4 for $c>0$. In fact, 
the Monte Carlo simulations performed prior to the 
present work (see \cite{ctbj,ajw} and in particular \cite{aa}) 
have not been precise enough to settle 
the question whether $d_h$ remains 4 for $c>0$ or 
whether it changes according to \rf{1.9} as a function of $c$.
In particular the most extensive simulations performed 
so far, rapported in \cite{aa}, were agonizing since
the results seemed to depend on the observable one used.
Finite size analysis using the spin correlators seemed 
to favor $d^W_h(c)$ although with somewhat large 
error bars, while the use of geometric quantity 
favored the $d_h=4$ hypothesis for $c\geq 0$.
Recently arguments have been given \cite{dupl} in favor of $d_h=4$.

In this article we will provide a new method for measuring $d_h$
based on 2d geometries with toroidal topology. The method allows
us to measure $d_h$ with high precision and leads 
to agreement with $d^W_h(c)$ given by \rf{1.9} also for $0<c<1$.
To be precise, we will show that $d_h^W(c)$ describes the 
fractal geometry for two unitary matter systems coupled
to 2d quantum gravity, namely the $c=1/2$ and $c=4/5$ 
conformal field theories. Until now formula \rf{1.9}
has only been verified convincingly for $c=0$ and (numerically) for the 
somewhat artificial, analytic continuation of \rf{1.1a}
to negative $D$.   

\section{Measuring the toroidal Hausdorff dimension}
\subsection{The set up}
As described above we expect geodesic distances to scale 
anomalously in 2d Euclidean quantum gravity. In \cite{ab3,ab1,ab2}
this was used to measure $d_h$ for spacetimes with toroidal topology.
The basic observation was that a shortest non-contractible 
loop (if it exists) is a geodesic curve. 
Let the intrinsic Hausdorff dimension be $d_h$,
let the spacetime volume be $V$ and let the probability distribution
of the shortest non-contractible loop be $P_V(L)$. Since $x= L/V^{1/d_h}$ is 
dimensionless and so is $P(L) dL$, we expect a relation
\beq\label{2.1}
P_V(L) = V^{-1/d_h} F(x),
\eeq
where the function $F(x)$ contains no dependence of $V$ except
the one found in $x$. From \rf{2.1} one obtains that $\la L \ra
\sim V^{1/d_h}$ and it was this scaling relation which was used 
in \cite{ab3,ab1,ab2} to determine $d_h$. Here we will directly use 
relation \rf{2.1} to determine $d_h$. 

If we regularize 2d quantum gravity using DT, the relation \rf{2.1}
can be read as a standard finite size scaling relation, using 
the terminology from the theory of critical phenomena. The volume $V$ is replaced by the 
number of triangles $N$ in the triangulation, a loop consists 
of links and $L$ is replaced by 
number of links $\ell$ in the shortest non-contractible loop for 
the given triangulation. Finally, the probability distribution $P_V(L)$
refers to the distribution obtained for the ensemble of toroidal geometries
for a quantum gravity theory with the  partition function \rf{1.7}. In
the regularized setting of DT the conformal field theory with 
central charge $c$ is represented in some way. As an example, the 
Ising model can be represented by Ising spins on  the triangles, interacting 
with the spins on the neighboring triangles, and the (inverse) temperature
$\b$ put equal to the critical value $\b_c$ to ensure the Ising model
represents a $c=1/2$ conformal field theory. In such a theory we can 
determine  $P_N(\ell)$  numerically by generating a large number of independent 
triangulations (with the weight of matter taken into account) and then 
simply measure the shortest non-contractible loop for each configuration.
Eq.\ \rf{2.1} is then replaced by the finite size scaling relation
\beq\label{2.2}
P_N(\ell) = N^{-1/d_h} F(x),~~~~x=\frac{\ell}{N^{1/d_h}}.
\eeq
By measuring $P_N(\ell)$ for different $N$ we will can easily 
test if the scaling assumption is satisfied and determine by 
standard fitting the best value of $d_h$. Of course 
we cannot expect \rf{2.2} to be valid and reproduce 
a continuum formula like \rf{2.1} for small values of $N$, so 
usually the fitting procedure will also involve discarding 
some range of small $N$.

The length of a shortest non-contractible loop is just the first of a sequence
of natural geodesic lengths, $0<\ell_1 \leq\ell_2 \leq \cdots$ that one can assign to a torus.
These $\ell_i$ are defined in the following way.
First one notices that the closed curves on a torus fall into homotopy classes.
In the piecewise linear geometries like the ones generated by DT, these classes contain the closed paths consisting of sequences of edges that can be deformed into eachother with local deformations. 
We call a homotopy class simple if it contains a simple curve, i.e.\
a curve which does not intersect itself.
In each non-trivial simple homotopy class $\Gamma$
we can find one or more paths which have a minimum length $\ell_{\Gamma}$ and which can regarded as discrete geodesics.
If we order the lengths $\ell_{\Gamma}$ for all $\Gamma$ we obtain a well-defined sequence of lengths
\beq\label{2.3}
0 < l_1 \leq l_2 \leq \cdots
\eeq
and a corresponding sequence of probability distributions:
\beq\label{2.4}
P_N^{(i)}(\ell_i) = N^{-1/d_h}F_i\Big(\ell_i/N^{1/d_h}\Big).
\eeq

It turns out to be advantageous to use $P_N^{(2)}(\ell_2)$ in the determination 
of $d_h$. In average $\ell_2$ will be larger than $\ell_1$ (see Fig.\ \rf{fig2} below) and the resulting 
probability distribution is significantly less sensitive to finite size 
effects. We do not fully understand why this is so,
but it makes it worth to construct $P_N^{(2)}(\ell_2)$, even if 
it is more computer-demanding to find $\ell_2$. 

\begin{figure}
\begin{center}
\includegraphics[width=7cm]{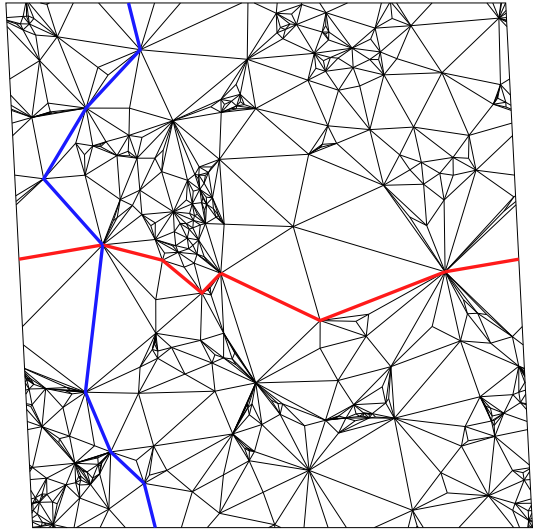}
\end{center}
\caption{A random triangulation of the torus coupled to Ising spins. 
The thick red curve is a shortest non-contractible loop of 
length $\ell_1=6$, while the thick 
blue curve is the ``second shortest'' 
loop of length $\ell_2=7$. The embedding of the triangulation in a
parallelogram in the complex plane is constructed using the methods in \cite{ab2}. }
\label{fig1}
\end{figure}

In \cite{ab1} we described an algorithm which allows one to 
find $\ell_1$ and the corresponding loop $\g_1$ 
efficiently for large triangulations. To find $\ell_2$ we are 
looking for a simple closed path $\g_2$ of minimal length 
which is neither homotopic to $\g_1$ nor to a point.  
This implies that it has  to intersect $\g_1$ at least once.
Therefore we can find $\gamma_2$ by performing for each vertex 
in $\gamma_1$ a search starting at that vertex in the following way.
First we inspect the neighbors of the starting vertex, 
then the neighbors of the neighbors, etc.
Encountering a vertex that has been visited before 
means that one has discovered a loop in the triangulation.
We stop at the first such loop which is neither homotopic to $\gamma_1$ 
nor to a point. It is not hard to see that the length of 
this loop is the sought-after length $\ell_2$.
Examples of the curves $\gamma_1$ and $\gamma_2$ for a 
large random triangulation $T$ are shown in Fig.\ \ref{fig1}.

\subsection{The numerical results}

We use the method described above to determine $d_h$ for $c=-2,0,1/2$ and 4/5.
For $c=-2$ and 0 we have already determined $d_h$ with good precision
measuring $\la\ell_1\ra$ corresponding to the distributions 
$P_N^{(1)}(\ell_1)$, 
so the determination of $d_h$ for these values 
of $c$ using finite size scaling directly on the distribution 
$P_N^{(2)}(\ell_2)$ is basically a check that the method works.\footnote{In fact, the distributions of closely related observables have been calculated analytically in \cite{guitter} for $c=0$. The author obtained expressions for the length of a shortest non-contractible loop passing via a marked point on a random bipartite quadrangulation of the torus and similarly for a second shortest loop, obtaining the expected scaling with $N^{1/4}$. } 
The triangulations are generated as described in \cite{ab1,ab2} 
by Monte Carlo simulations for $c=0$
and by a recursive algorithm for $c=-2$. The class of triangulations
used is the most general one where links in a triangle are allowed 
to be identified and where pairs of triangles are  allowed to share 
more than one edge.  The $c=1/2$ system is realized as Ising spins 
on the triangles, interacting with spins on neighboring triangles. 
The $c=4/5$ system is realized as a 3-states Potts model, the spins
also on the triangles. For a fixed triangulation the partition function 
takes the form
\beq
Z_Q = \sum_{\sigma\in\{1,\ldots,Q\}^N} \exp\left(2\beta_Q\sum_{\la ij\ra} \delta_{\sigma_i\sigma_j}\right),
\eeq 
with $Q=2$ for the Ising spins and $Q=3$ for the 3-states Potts model. The sum in the exponential is over the pairs of adjacent triangles $i$ and $j$.
The coupling constants $\beta_Q$ are chosen to be the critical coupling in the infinite volume limit (i.e. the $N \to \infty$ limit). For the general ensemble of triangulations
used here, and with the spins on the triangles these are known: 
$\beta_2=1/2\log(1+2\sqrt{7}) \approx 0.9196$ and $\beta_3=1/2\log(1+\sqrt{47}) \approx 1.0306$ (see \cite{crit-beta} and \cite{daul,ctbj} respectively). The updating of the triangulations is done in the standard way
by flipping the diagonal of a randomly selected pair of adjacent triangles.
The Wolff algorithm \cite{wolff} is used to efficiently update the spins at the 
critical points.

\begin{figure}
\begin{center}
\includegraphics[width=10cm]{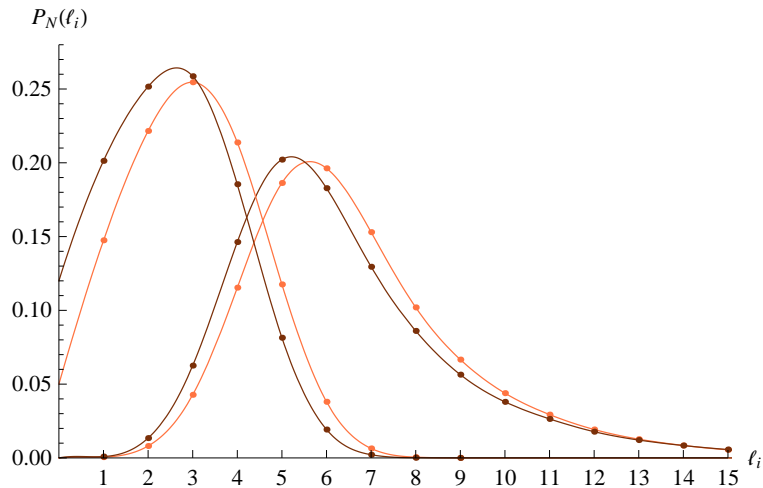}
\end{center}
\caption{The probability distributions $P_N^{(i)}(\ell_i)$, $i=1,2$,
for dynamical triangulations coupled to the Ising model (light points) and
coupled to the Potts model (dark points). The curves are smooth interpolations 
used for finding the scaling. The left curves are $P_N^{(1)}(\ell_1)$  and 
the right curves are  $P_N^{(2)}(\ell_2)$, all for $N=8000$}
\label{fig2}
\end{figure}

The simulations were performed with triangulations of the torus 
with $N=250$ up to $N=256000$ triangles.
First the systems were thermalized by performing roughly $10^5$ sweeps, 
where a sweep corresponds to on average $N$ random update moves on the 
triangulation and $N$ spin flips.
After thermalization between $10^5$ and $10^6$ measurements of $\ell_1$ 
and $\ell_2$ were performed with 100 sweeps in between measurements. 
From the measurements one can construct the probability distribution 
$P_N^{(i)}(\ell_i)$ for obtaining a particular length $\ell_i$ 
for a random triangulation with $N$ triangles, 
as shown in figure \ref{fig2} for $N=8000$.

\begin{figure}[t]
\begin{center}
\includegraphics[width=14cm]{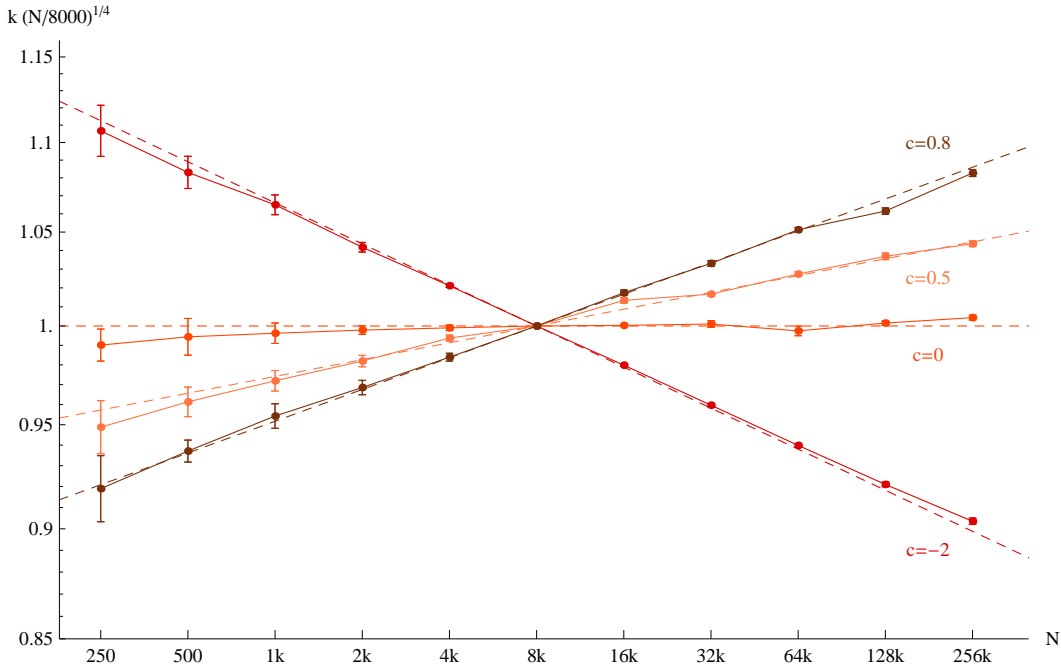}
\end{center}
\caption{The values $k$ leading to a best fit $P_N^{(2)}(\ell_2) 
\propto f(k \ell_2)$, where $f(\tilde{\ell}_2)$ is the smooth 
interpolation of $P_{N=8000}^{(2)}(\ell_2)$ shown in figure \ref{fig2}. 
The shown errors are predominantly systematic, as a result of the 
fact that the profiles $P_N^{(2)}(\ell_2)$ do not exactly collapse. 
The dash lines correspond to $k \propto N^{-1/d_h^W}$ with $d_h^W$ 
the values given by formula \rf{1.9}.  }
\label{fig3}
\end{figure}

To study the finite size scaling we fit 
$P_N^{(2)}(\ell_2) \propto f( k \ell_2)$, where $f(\tilde{\ell}_2)$ 
is a smooth interpolation of the distribution $P_{N=8000}^{(2)}(\ell_2)$ 
obtained for $N=8000$ triangles. 
When performing the fit for a fixed $N$ we took into account only the data points for which $P_N(\ell_2)$ was larger than $0.1$ times the largest $P_N(\ell_2)$.
We made this choice based on the fact that the collapse of the $P_N(\ell_2)$ for different $N$ according to (\ref{2.2}) is very accurate for the peak of the distribution but less so for its tail at large $\ell_2$.
One may view the values of $k$ obtained by such a best fit to be an accurate measurement of (the inverse of) the position of the peak in the distribution $P_N(\ell_2)$.
The results of the best fits for $k$ as function of $N$ are shown in figure \ref{fig3}.
We have rescaled $k$ by $(N/8000)^{1/4}$, 
such that a flat curve would correspond to $k \propto N^{-1/4}$. 
By construction the points for $N=8000$ take the value $1$.
On the figure we have also shown the data for $c=0$ and $c=-2$ as 
described above.

\begin{figure}
\begin{center}
\includegraphics[width=10cm]{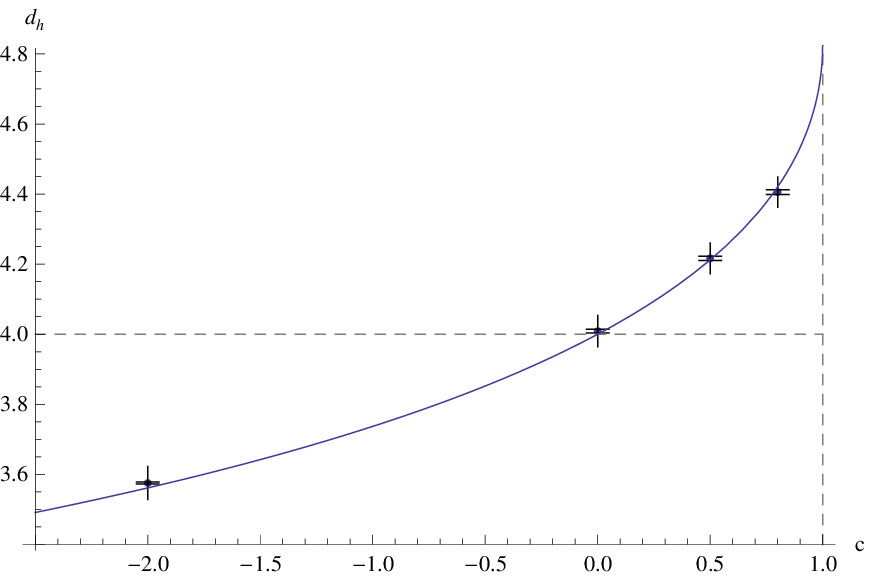}
\end{center}
\caption{The values for the Hausdorff dimension $d_h$ as extracted 
from the slope of the curves in figure \ref{fig3}. 
The curve corresponds to formula \rf{1.9}.}
\label{fig4}
\end{figure}

The fitted values $k$ show a beautiful scaling with the 
volume $N$ and are in agreement with $k\propto N^{-1/d_h^W}$ with $d_h^W$ 
given by formula \rf{1.9}, as indicated by the dash lines. 
Extracting the slopes by a fit one obtains the values in 
the following table, which are also plotted in figure \ref{fig4}.

\begin{center}
\begin{tabular}{r|c|c}
$c$ & $d_h$ (by fit) & $d_h$ (theoretical) \\
\hline 
$-2$ & $3.575 \pm 0.003$ & $3.562$ \\
$0$ & $4.009 \pm 0.005$ & $4.000$ \\
$1/2$ & $4.217 \pm 0.006$ & $4.212$ \\
$4/5$ & $4.406 \pm 0.007$ & $4.421$ \\
\hline
\end{tabular}
\end{center}

\section{Discussion}

As seen from the table above and Fig.\ \ref{fig4} 
the agreement between formula \rf{1.9} and
the values of $d_h(c)$ is impressive, in particular when one compares
to the precision one could obtain using the older methods for extracting 
$d_h$. Our present measurements provide convincing numerical evidence 
that the peak in the distribution of second shortest loop 
lengths scales according to a Hausdorff dimension given by the 
Watabiki formula \rf{1.9}. It is interesting to compare 
in more detail with results obtained in \cite{aa}, where the 
Hausdorff dimension was extracted for spacetimes with spherical 
topologies, using two-point correlation functions, either 
entirely geometric or spin-spin correlation functions. Finite 
size scaling analysis applied to the spin-spin correlation 
functions the same way as finite size scaling has been applied 
in this article led to agreement with $d_h^W(c)$, although 
with large errorbars. However, the finite size scaling applied to 
geometric correlators and expectation values, as well as the 
study of  the short distance part of 
the correlators, spin or geometric, led to 
better agreement with $d_h(c) =4$ for $0 \leq c< 1$.

In this article we have only been looking at geometric quantities
and we have for the first time seen that they scale according 
to $d_h^W(c)$ for $c>0$, even quite convincingly, with 
significantly smaller error bars than were 
present in measurement of geometric corelators in the work \cite{aa}. 
Does that rule out the  $d_h(c) =4$ hypothesis for $0\leq d_h  \leq 1/4$? 
Not entirely. As mentioned above, we 
obtained the clean scaling results by cutting away the tales
of the distributions  $P_{N}^{(2)}(\ell_2)$ (we imposed 
$P_N^{(2)}(\ell_2)\geq 0.1 \, \max (P_N^{(2)}(\ell_2))$).  
In fact, preliminary investigation of the tail of 
the distribution has shown that the decay of $P_{N}^{(2)}(\ell_2)$ 
for large $\ell_2$ is approximately exponential 
and that the decay rate scales more like $N^{-1/4}$ than $N^{-1/d_h^W}$. 
However, more statistics and larger system sizes are 
required to decide the matter.
 
\vspace{.5cm}      
\noindent {\bf Acknowledgments.} 
The authors acknowledge support from the ERC-Advance grant 291092,
``Exploring the Quantum Universe'' (EQU). JA acknowledges support 
of FNU, the Free Danish Research Council, from the grant 
``quantum gravity and the role of black holes''. Finally this research 
was supported in part by the Perimeter Institute of Theoretical Physics.
Research at Perimeter Institute is supported by the Government of Canada
through Industry Canada and by the Province of Ontario through the 
Ministry of Economic Development \& Innovation.


\begin{thebibliography}{99}

\bibitem{book}
  J.~Ambjorn, B.~Durhuus and T.~Jonsson,
  Cambridge, UK: Univ. Pr., 1997. (Cambridge Monographs in Mathematical Physics). 363 p





\bibitem{spectrald}
  J.~Ambjorn, D.~Boulatov, J.~L.~Nielsen, J.~Rolf and Y.~Watabiki,
  JHEP {\bf 9802} (1998) 010
  [hep-th/9801099].

\bibitem{watabiki}
  Y.~Watabiki,
  Prog.\ Theor.\ Phys.\ Suppl.\  {114} (1993)  1-17.


\bibitem{dk}
  J.~Distler, Z.~Hlousek, H.~Kawai,
  Int.\ J.\ Mod.\ Phys.\  A\ 5 (1990)  1093.




\bibitem{transferm}
H.~Kawai, N.~Kawamoto, T.~Mogami, Y.~Watabiki,
Phys.\ Lett.\ B\ 306 (1993) 19-26
[hep-th/9302133].


\bibitem{aw}
J.~Ambj\o rn, Y.~Watabiki,
Nucl.\ Phys.\ B\ 445 (1995) 129-144 [hep-th/9501049].


\bibitem{kazakov}
  V.~A.~Kazakov,
  Phys.\ Lett.\  {A\ 119} (1986)  140-144.
  




\bibitem{ki}
  N.~Ishibashi and H.~Kawai,
  Phys.\ Lett.\ B {\bf 322} (1994) 67
  [hep-th/9312047].


  
\bibitem{aajk}
  J.~Ambj\o rn, K.N.~Anagnostopoulos, J.~Jurkiewicz, C.F.~Kristjansen,
  JHEP {9804} (1998)  016
  [hep-th/9802020].


\bibitem{kas}
  N.~Kawamoto, V.A.~Kazakov, Y.~Saeki, Y.~Watabiki,
  Phys.\ Rev.\ Lett.\  {68} (1992)  2113-2116.


  
\bibitem{dminustwo}
  J.~Ambj\o rn, K.N.~Anagnostopoulos, T.~Ichihara, L.~Jensen, N.~Kawamoto, Y.~Watabiki, K.~Yotsuji,
  Phys.\ Lett.\  {B\ 397} (1997)  177-184
  [hep-lat/9611032];
  Nucl.\ Phys.\  {B\ 511} (1998)  673-710
  [hep-lat/9706009].


\bibitem{ab3}
  J.~Ambjorn and T.~G.~Budd,
  Phys.\ Lett.\ B {\bf 718} (2012) 200
  [arXiv:1209.6031 [hep-th]].


\bibitem{ab1}
  J.~Ambjorn, J.~Barkley, T.~Budd and R.~Loll,
  Phys.\ Lett.\ B {\bf 706} (2011) 86
  [arXiv:1110.3998 [hep-th]].

\bibitem{ab2}
  J.~Ambjorn, J.~Barkley and T.~G.~Budd,
  Nucl.\ Phys.\ B {\bf 858} (2012) 267
  [arXiv:1110.4649 [hep-th]].



\bibitem{dupl}
  B.~Duplantier,
  arXiv:1108.3327 [math-ph].

\bibitem{crit-beta}
  D.~V.~Boulatov, V.~A.~Kazakov,
  Phys.\ Lett.\ B {\bf 186} (1987) 379.
  
\bibitem{daul}
  J.-M.~Daul, hep-th/9502014.
  
\bibitem{ctbj}
  S.~Catterall, G.~Thorleifsson, M.~Bowick, V.~John,
  Phys.\ Lett.\ B {\bf 354} (1995) 58 [hep-lat/9504009].

\bibitem{ajw}
  J.~Ambjorn, J.~Jurkiewicz, Y.~Watabiki,
  Nucl.\ Phys.\ B {\bf 454} (1995) 313 [hep-lat/9507014].

\bibitem{aa}
  J.~Ambjorn and K.~N.~Anagnostopoulos,
  Nucl.\ Phys.\ B {\bf 497} (1997) 445
  [hep-lat/9701006].


\bibitem{guitter}
  E.~Guitter,
  J.\ Stat.\ Mech.\ (2010) P04018 
  [arXiv:1003.0372 [math-ph]].
  
\bibitem{jonsson}
  T.~Jonsson,
  Phys.\ Lett.\ B {\bf 425} (1998) 265
  [hep-th/9801150].  
  
\bibitem{wolff}
  U.~Wolff,
  Phys.\ Rev.\ Lett. 62 (1989) 361.
  



\end{thebibliography}
\end{document}